\documentclass[aps,pra,twocolumn,showpacs,groupedaddress,superscriptaddress,nofootinbib]{revtex4}
\usepackage{bbm,amsmath,amssymb,amsbsy,graphicx,subfigure,txfonts}

\usepackage{hyperref,color}
\usepackage[latin1]{inputenc}
\newcommand{\be}{\begin{equation}}
\newcommand{\ee}{\end{equation}}
\newcommand{\ben}{\begin{eqnarray}}
\newcommand{\een}{\end{eqnarray}}
\newcommand{\bes}{\begin{subequations}}
\newcommand{\ees}{\end{subequations}}

\newcommand{\T}{^{\sf T}}

\newcommand{\tr}{\text{tr}}
\newcommand{\gr}[1]{\boldsymbol{#1}}
\newcommand{\ket}[1]{|#1\rangle}
\newcommand{\bra}[1]{\langle#1|}

\newcommand{\sig}{{\gr\sigma}}
\newcommand{\gam}{\boldsymbol{\gamma}}

\begin{document}
\title{Theory of genuine tripartite nonlocality of Gaussian states}

\date{October 9, 2013}
\author{Gerardo Adesso}
%\email{gerardo.adesso@nottingham.ac.uk}
\affiliation{School of Mathematical Sciences, The University of Nottingham,
University Park, Nottingham NG7 2RD, United Kingdom}

\author{Samanta Piano}
\affiliation{Midlands Ultracold Atom Research Centre, School of Physics and Astronomy, The University of
Nottingham, Nottingham NG7 2RD, United Kingdom}

\begin{abstract}
We investigate the genuine multipartite nonlocality of three-mode Gaussian states of continuous variable systems. For pure states, we present a simplified procedure to obtain the maximum violation of the Svetlichny inequality based on displaced parity measurements, and we analyze its interplay with genuine tripartite entanglement measured via R\'enyi-$2$ entropy. The maximum Svetlichny violation admits tight upper and lower bounds at fixed tripartite entanglement. For mixed states, no violation is possible when the purity falls below $0.86$. We also explore a set of recently derived weaker inequalities for three-way nonlocality, finding violations for all tested pure states. Our results provide a strong signature for the nonclassical and nonlocal nature of Gaussian states despite their positive Wigner function, and lead to precise recipes for its experimental verification.
\end{abstract}


\maketitle

Quantum mechanics defies our intuition every day. At the level of a single system, with superposition phenomena \cite{nv}. At the level of composite systems, with correlations which are incompatible with the classical notion of locality \cite{EPR35,bell,chsh}. In a hierarchy of possible manifestations of the quantum fabric of the world, nonlocality is perhaps the strongest one. States violating a Bell-type inequality \cite{bellrev} are necessarily entangled \cite{entanglement}, therefore marking their departure from a classical description \cite{schr}, while the converse is not always true \cite{werner89}. In recent years, nonlocality has come to be appreciated as an operational resource \cite{operationonloc}, which plays an essential role e.g.~for the implementation of secure quantum key distribution \cite{qkd}. The concept of multipartite nonlocality is however significantly less understood than its bipartite counterpart \cite{bell,chsh,merminklyshko,svelt,bancalprl,sveltghzw,expsvelt,othermulti,operationonloc,bancal,bellrev}. For a tripartite system, the conventionally employed inequality, whose violation signals a genuine three-way nonlocality, is due to Svetlichny \cite{svelt}. Such an inequality can be violated e.g.~by both Greenberger-Horne-Zeilinger (GHZ) and $W$ classes of states for three qubits \cite{sveltghzw,expsvelt}, and is essentially {\it the} unique witness of nonlocality when all three parties perform two measurements with two outcomes each \cite{bancal}. More recently, a reassessment of tripartite nonlocality has led to a series of weaker inequalities, whose violation can reveal tripartite nonlocality even when the Svetlichny one is not violated \cite{operationonloc,bancal}. According to this framework, it has been conjectured that all fully inseparable tripartite pure states are three-way nonlocal \cite{bancal}. In general, the interplay between the quantitative violation of these inequalities and the degree of genuine tripartite entanglement in the tested states remains unclear \cite{sveltghzw}. We recall that multipartite entanglement, i.e.~entanglement encoded in systems of more than two parties, can arise in many inequivalent forms \cite{entanglement}. In qubit systems, e.g., GHZ and $W$ states are both fully inseparable, but in GHZ states only genuinely tripartite entanglement is present, while $W$ states have maximum bipartite entanglement between any pair of qubits. Generally, multipartite entanglement obeys strict monogamy constraints on its distribution \cite{ckw}.

Nonlocality tests have been studied also for continuous variable systems \cite{homobell,kb,jeong,bramermin,chen,ferraroparis,maurorefs,jie,mauro}, namely systems whose canonical degrees of freedom, which can be nonclassically correlated, have a continuous spectrum \cite{brareview,gaussreview}. This is the case, for instance, for quadrature modes of light, phononic momentum modes of Bose-Einstein condensates, vibrational modes of mechanical resonators, or collective spin components of cold atomic ensembles \cite{book}. In typical studies of continuous variable nonlocality, measurements on these systems are binned to return a dichotomic result, so that the traditional format of Bell-type inequalities can be straightforwardly adopted \cite{homobell}. Continuous variable systems constitute a powerful setting for the unconditional demonstration of quantum communication and cryptography protocols, often outperforming their qubit counterparts thanks to high quality feasible measurements, despite the unavailability of arbitrarily large entanglement \cite{brareview}. A central class of continuous variable states is represented by Gaussian states, which include coherent, squeezed and thermal states of harmonic Hamiltonians \cite{ourreview,gaussreview,book}. Notwithstanding their limitations \cite{nogo}, Gaussian states and operations are preferred resources which can be realized in a plethora of experimental platforms \cite{book}. On the theoretical side, equally, considerable effort has been devoted to characterize their informational properties and correlation structure \cite{ourreview,gaussreview,renyi}.
%Very recently, for instance, a consistent theory of their bipartite and multipartite correlations (classical, quantum, and total) has been achieved adopting the R\'enyi entropy of order $2$ as unit of information \cite{renyi}.

To the best of our knowledge, nevertheless, no systematic study of multipartite nonlocality of Gaussian states has been reported so far: the purpose of the present paper is to fill this gap. Investigating the nonlocal character of Gaussian states is of special importance, as according to some criteria Gaussian states are essentially classical, being associated to a positive Wigner distribution in phase space \cite{nonclasswig}. This is often seen as an argument against their use as genuine quantum resources, since computations involving Gaussian states can be simulated classically \cite{gaussimul}. Similarly, Bell tests involving homodyne detections cannot be violated by Gaussian states \cite{homobell}. However, a breakthrough came when displaced parity measurements (which are non-Gaussian operations)  were considered to design continuous variable nonlocality tests \cite{kb}. The valuable feature of these measurements is that their expectation value is directly proportional to the Wigner distribution of the state at a given phase space point \cite{kb}. Using these measurements, bipartite nonlocality of Gaussian states was predicted theoretically  \cite{kb,jeong,bramermin}. Recently, a prescription to extend the Svetlichny inequality to the continuous variable domain using these measurements has been proposed \cite{mauro}. This formalism was however applied in the Gaussian setting only to sparse, specialized examples \cite{bramermin,jie,mauro,entropy}.

Here we perform a comprehensive analysis of the phase space Svetlichny inequality  \cite{svelt,bancalprl,mauro} for three-mode Gaussian states \cite{3mpra,3mnjp}.  We provide analytical prescriptions for detecting the maximum violation of the Svetlichny inequality for symmetric pure states. For generic nonsymmetric mixed states, we  investigate numerically the interplay between tripartite nonlocality, tripartite entanglement \cite{renyi}, and state purity. We further consider alternative tests of tripartite nonlocality which exclude signalling correlations \cite{bancal}, and we find violations for all tested fully inseparable pure states. These results advance substantially our understanding of entanglement and nonlocality in multipartite infinite-dimensional systems.
%are presented in the following, after a recap on Gaussian states and phase space nonlocality tests.

\noindent{\bf Three-mode Gaussian states.}---
We consider a $n$-mode continuous variable system; we collect the quadrature operators in the vector $\underline{\hat R} = (\hat q_1, \hat p_1, \hat q_2, \hat p_2, \ldots, \hat q_n, \hat p_n)^{\sf T} \in \mathbb{R}^{2n}$ and write the canonical commutation relations compactly as $[\hat R_j, \hat R_k] = i \left(\boldsymbol\omega^{\oplus n}\right)_{j,k}$ with $\boldsymbol\omega = {{\ 0 \ \ 1} \choose {-1 \ 0}}$ being the symplectic matrix. A Gaussian state $\rho$ is represented by a positive, Gaussian   Wigner distribution in phase space,
\begin{equation}
\label{wigner}
W_\rho(\underline{\xi}) = \frac{1}{\pi^n \sqrt{\det{\sig}}} \exp\big(-\underline{\xi}^{\sf T} \sig^{-1} \underline{\xi}\big)\,,
\end{equation}
where $\underline{\xi} \in \mathbb{R}^{2n}$, and $\sig$ is the covariance matrix (CM) of the second moments $\sigma_{j,k}=\text{tr}[\rho \{\hat{R}_j,\hat{R}_k\}_+]$, which completely characterizes the  state $\rho$ up to local displacements \cite{ourreview}. The purity of the state is $\mu(\rho) = \tr\,\rho^2 = (\det\sig)^{-\frac12}$.

For a three-mode  ($n=3$)  state $\rho$, we have $\underline\xi =(\underline \xi_1, \underline \xi_2, \underline \xi_3) =(q_1,p_1,q_2,p_2,q_3,p_3)$, and $\sig \equiv \sig_{123}$ given in block form by
\begin{equation}
\label{sigma}
\sig=\left(
\begin{array}{c|c|c}
\sig_1 & \gam_{12} & \gam_{13} \\ \hline
\gam_{12}\T & \sig_2 & \gam_{23} \\ \hline
\gam_{13}\T & \gam_{23}\T & \sig_3
\end{array}
\right)\,.
\end{equation}
A {\it pure} three-mode Gaussian state can be transformed by local unitaries in a standard form \cite{3mpra} characterized by $\sig_j = \text{diag}\{a_j,a_j\}$, $\gam_{jk} = \text{diag}\{g_{jk}^+,g_{jk}^-\}$, with $g_{jk}^\pm = \left\{\sqrt{[(a_i-1)^2-(a_j-a_k)^2][(a_i+1)^2-(a_j-a_k)^2]}\right.\pm\left.\sqrt{[(a_i-1)^2-(a_j+a_k)^2][(a_i+1)^2-(a_j+a_k)^2]}\right\}/\big[4\sqrt{a_j a_k}\big]$, where $|a_j - a_k|+1\leq a_i\leq a_j+a_k-1$ and  $\{i,j,k\}$ is a permutation of $\{1,2,3\}$. Any pure state with $a_i>1$ $\forall i=1,2,3$ is fully inseparable and its genuine tripartite entanglement, as emerging from the monogamy inequality \cite{ckw,contangle,3mpra,renyi}, can be quantified by the residual R\'enyi-$2$ entanglement entropy ${\cal E}_{1|2|3}(\sig)$, whose explicit expression in terms of $a_{1,2,3}$ is reported in \cite{renyi}.
For {\it mixed} three-mode Gaussian states, there are more layers of separability, as classified in \cite{giedke}; see also   \cite{3mpra,3mnjp} for further details.

\begin{figure*}[tbh]
\begin{minipage}[b]{5.4cm}
\centering
\subfigure[]{\includegraphics[width=5.4cm]{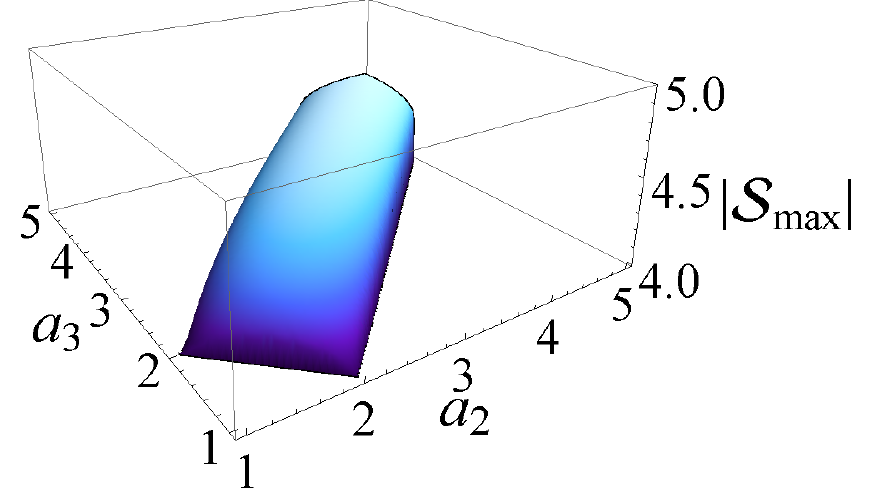}}
\subfigure[]{\includegraphics[width=5.4cm]{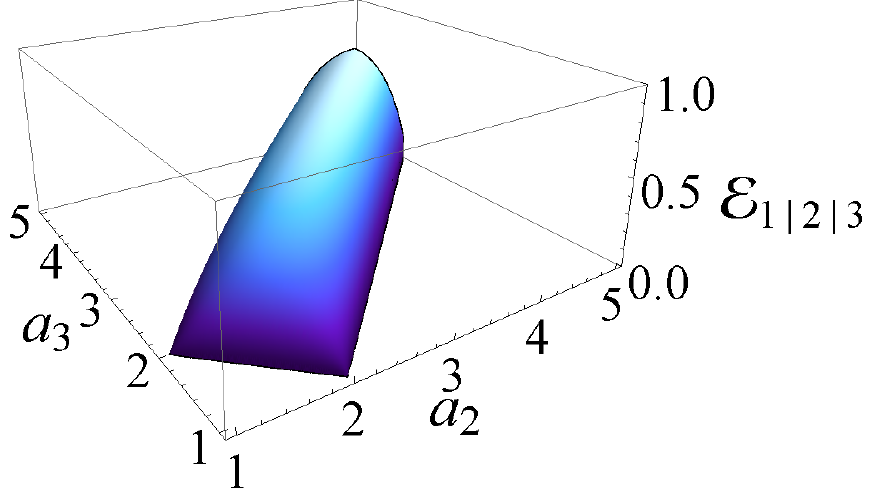}}\\
\end{minipage}
\hspace{0.2cm}
\begin{minipage}[b]{5.2cm}
\centering
\subfigure[]{\includegraphics[width=5.1cm]{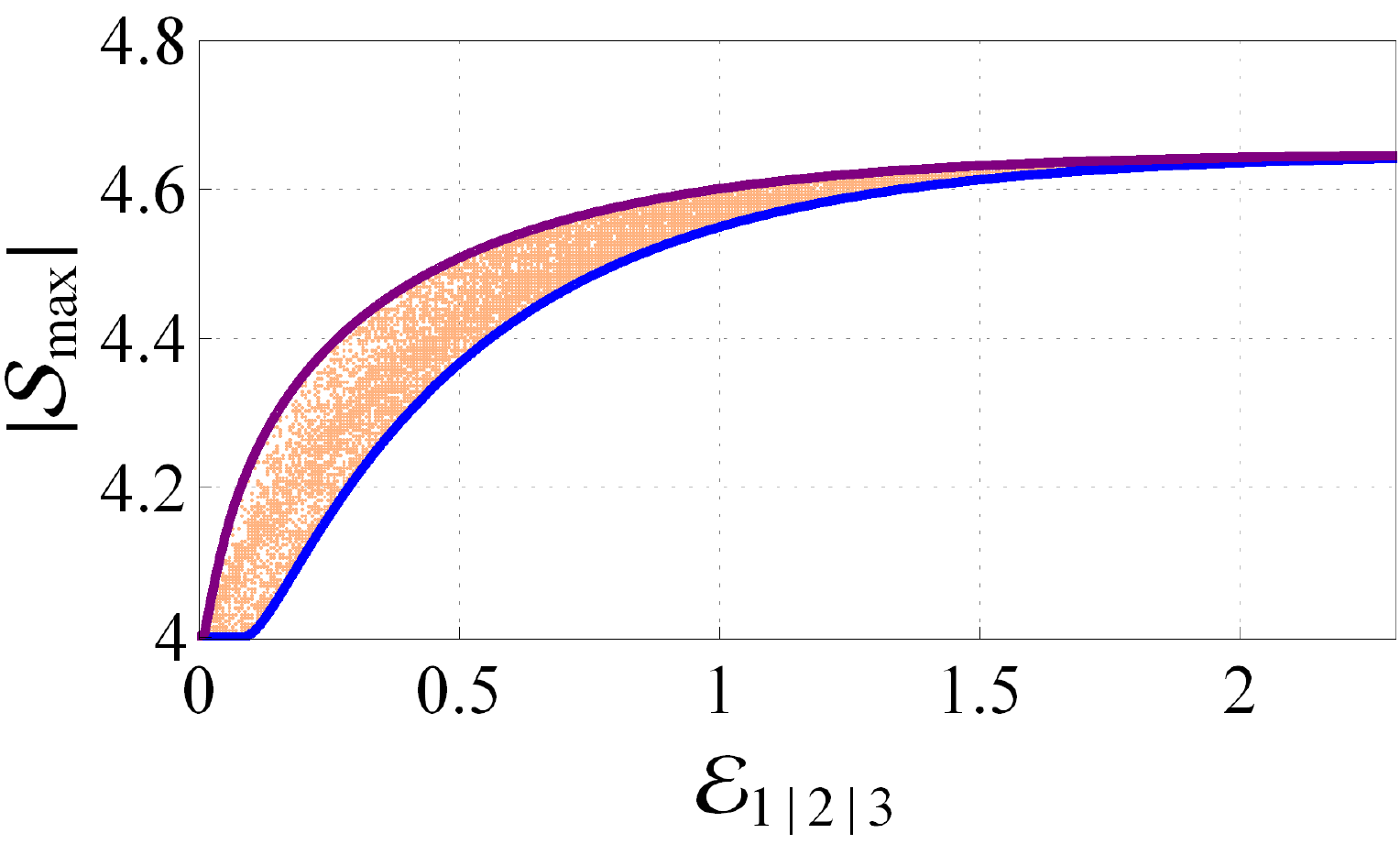}} \\
\subfigure[]{\includegraphics[width=5.25cm]{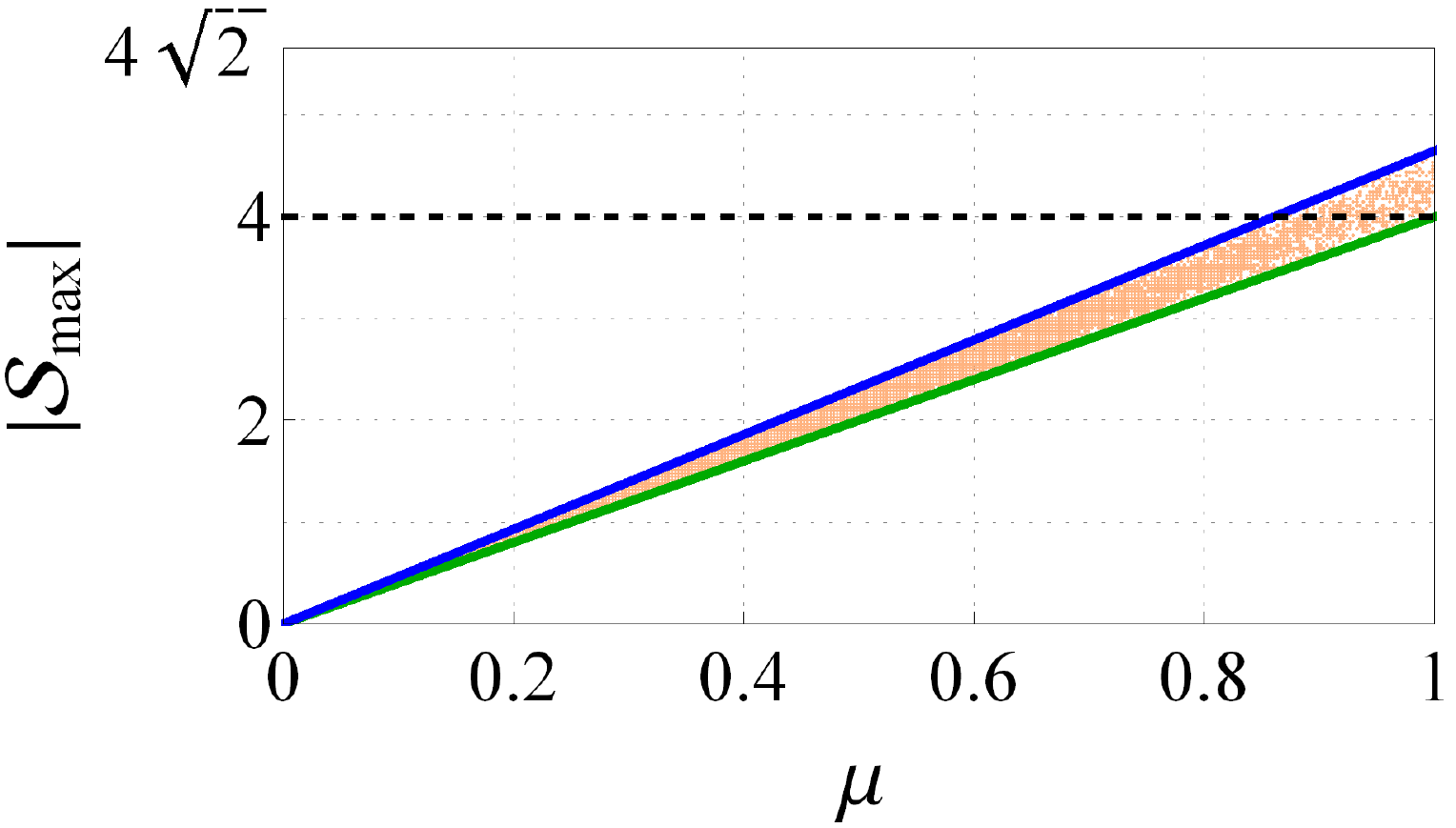}}
\end{minipage}
\hspace{0.2cm}
\begin{minipage}[b]{6.5cm}
\centering
\subfigure[]{\includegraphics[height=6.5cm]{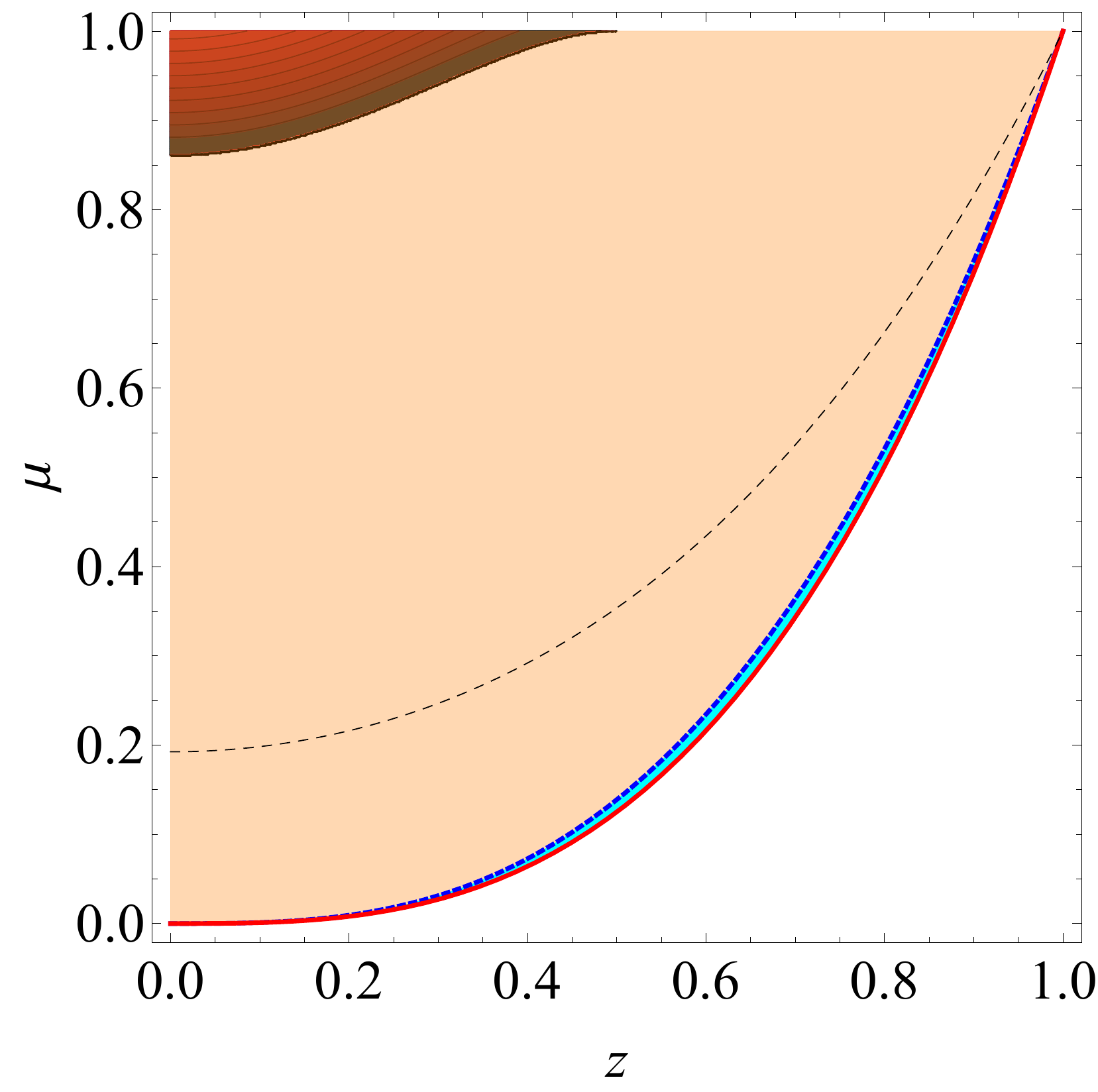}}
\end{minipage}
\caption{Plots of (a) maximum Svetlichny parameter $|{\cal S}_{\max}|$ and (b) genuine tripartite R\'enyi-$2$ entanglement ${\cal E}_{1|2|3}$ for pure three-mode Gaussian states with $a_1=2$. Panel (c) shows a direct comparison between these two quantities for $10^5$ random pure three-mode Gaussian states. Panel (d) depicts the maximum Svetlichny parameter versus the state purity $\mu$ for $10^5$ random mixed three-mode Gaussian states. Panel (e) is a classification of the inseparability and Svetlichny nonlocality properties of fully symmetric mixed three-mode Gaussian states with CM $\sig^{ms}(a,\mu)$ in terms of their purity $\mu$ and of an inverse squeezing parameter $z=\frac{1}{2} \sqrt{12 a^2-3 f_a-8}$. Details of the boundary curves and regions for the various panels are provided in the main text. All the quantities plotted are dimensionless.
}
\label{figunghie}
\end{figure*}
%\begin{figure}[h]

\noindent{\bf Phase space Svetlichny inequality.}---
Quantum nonlocality can be revealed on composite systems by letting every party perform a selection of different measurements on their subsystems, each with two or more possible outcomes. Typically, a correlation function is constructed from the expectation values of the measured observables, whose value is bounded if a   hidden variable theory is assumed. Correlations exceeding the bound signal the failure of local realism and the correctness of the quantum description of the world. We refer the reader to \cite{bell,chsh,operationonloc,bancal,bellrev} and references therein for an updated account on the subject. In the case of a three-party system, when each party can perform two possible two-outcome measurements, Svetlichny derived an inequality whose violation reveals {\it genuine} tripartite nonlocality \cite{svelt}, in the sense that no  hybrid local-nonlocal hidden variable model can be compatible with the measured data when the inequality is violated. This inequality builds upon bipartite Bell-CHSH and multipartite Mermin-Klyshko inequalities \cite{bell,chsh,merminklyshko} and can be formulated as follows. Suppose Alice, Bob and Charlie share a tripartite state $\rho$, and each party $j=1,2,3$ measures the operator $P_j$ in two possible settings, $\xi_j$ and $\xi'_j$, where $\{P_j\}$ are operators with eigenvalues $\pm 1$. Then the Svetlichny parameter can be written as \cite{svelt,bancalprl} (other expressions can be obtained by relabelling the measurement outcomes)
\begin{eqnarray}\label{sveltpar}
&&{\cal S}(\rho;\xi_1,\xi_2,\xi_3,\xi'_1,\xi'_2,\xi'_3) \\
&&\quad=\big\langle P_1(\xi'_1) P_2(\xi_2) P_3(\xi_3)  + P_1(\xi_1) P_2(\xi'_2) P_3(\xi_3) \nonumber \\
 &&\qquad+ P_1(\xi_1) P_2(\xi_2) P_3(\xi'_3) - P_1(\xi'_1) P_2(\xi'_2) P_3(\xi'_3)  \nonumber \\
&&\qquad+P_1(\xi_1) P_2(\xi'_2) P_3(\xi'_3)  + P_1(\xi'_1) P_2(\xi_2) P_3(\xi'_3) \nonumber \\
&&\qquad+ P_1(\xi'_1) P_2(\xi'_2) P_3(\xi_3) - P_1(\xi_1) P_2(\xi_2) P_3(\xi_3)\big\rangle_\rho \nonumber \,.
\end{eqnarray}
Assuming hidden variables for the measurements of any two parties one gets the Svetlichny inequality \cite{svelt}
\begin{equation}\label{sveltina}
|{\cal S}| \leq 4\,,
\end{equation}
which can be instead violated quantum mechanically up to the value $|{\cal S}|=4\sqrt{2} \approx 5.65$, reached for instance on three-qubit GHZ states \cite{svelt,sveltghzw,expsvelt}. Any violation of inequality~(\ref{sveltina}) is a signature of genuine tripartite nonlocality on the state $\rho$, and implies that $\rho$ is fully inseparable. Typically, one needs to optimize the measurement settings to single out the maximum violation achievable on a given state, $|{\cal S}_{\max}(\rho)| = \sup_{\{\xi_j\},\{\xi'_j\}} |{\cal S}(\rho;\{\xi_j\},\{\xi'_j\})|$.

In continuous variable systems, we choose displaced parity measurements as the operators to be measured on each mode $j$ \cite{kb,jeong,mauro}. We have
$P_j(\underline \xi_j) = \sum_{n=0}^{\infty} (-1)^n \ket{\underline \xi_j,n}\bra{\underline \xi_j,n}$, where $\ket{\underline \xi_j,n}$ is the $n^{\rm th}$ Fock state displaced by $\underline \xi_j = (q_j, p_j)$. This operator can be measured by photon counting preceded by a phase space displacement, the latter implemented e.g. by beamsplitting the input mode with a tunable coherent field \cite{kb}. By definition of the Wigner function, one has  $\langle P_j(\underline \xi_j) \rangle_{\rho_j} = \pi W_{\rho_j}(\underline \xi_j)$ for an arbitrary single-mode state $\rho_j$. We can then redefine all the terms in Eq.~(\ref{sveltpar}) in terms of expectation values of the Wigner function of the three-mode state $\rho$ evaluated at  suitable phase space points. For instance, the first term reads $\big\langle P_1(\xi'_1) P_2(\xi_2) P_3(\xi_3)\big\rangle_\rho = \pi^3 W_\rho(\underline \xi'_1, \underline \xi_2, \underline \xi_3)$, and so on \cite{mauro}. In the case of $\rho$ being an undisplaced three-mode Gaussian state, whose Wigner function is given by Eq.~(\ref{wigner}), then the Svetlichny parameter depends on the elements of the CM $\sig$, as well as on the measurement settings $\{\underline \xi_j\}, \{\underline \xi'_j\}$.

\noindent{\bf Svetlichny nonlocality of Gaussian states.}---
%We note that the maximum violation of the Svetlichny inequality is independent of local unitary operations. Therefore we can focus on Gaussian states whose CMs are in their simplest possible standard form, and whose first moments are set to zero as announced earlier.
We begin by investigating pure fully symmetric states \cite{network}, with CM $\sig^s(a)$ given by Eq.~(\ref{sigma}) with $a_1=a_2=a_3 \equiv a$. In this case, the maximum of Eq.~(\ref{sveltpar}) can be found analytically, and is attained for phase space points $\xi_j=(0,p^\ast)$, $\xi'_j=(0,-p^\ast)$ $\forall\,j=1,2,3$, with $p^\ast=0$ if $a \leq \sqrt{3/2}$ and $p^\ast=\sqrt{(a/f_a) \tanh^{-1}[f_a/(4 a^2)]}$ otherwise, where $f_a=a^2-1+\sqrt{9a^4-10a^2+1}$.
The maximum Svetlichny parameter then becomes
\begin{equation}\label{sveltsym}
|{\cal S}_{\max}\big(\sig^s(a)\big)| = \left\{
                      \begin{array}{ll}
                        4, & \hbox{$a \leq \sqrt{\frac32}$;} \\
                        \frac{4 \left(4 a^2+3 f_a-4\right) \left(8 a^2-2 f_a-5\right){}^{\frac{3}{-8 a^2+2 f_a+8}} }{4 a^2+5}, & \hbox{otherwise.}
                      \end{array}
                    \right.
\end{equation}
Interestingly, despite these states being fully inseparable as soon as $a>1$ \cite{giedke}, the Svetlichny inequality (\ref{sveltina}) is only violated for  $a>\sqrt{3/2} \approx 1.22$.
%This is pretty close to the bound $a>5\sqrt{2}/6 \approx 1.18$ required for the use of these fully symmetric states as resources for the efficient implementation of Byzantine agreement with detectable broadcast \cite{byz}; this finding suggests that three-way Svetlichny nonlocality, more than three-partite entanglement, is the ingredient needed for such a multipartite cryptographic primitive, somehow mirroring the case of bipartite protocols \cite{qkd}.
The value $|{\cal S}_{\max}\big(\sig(a)\big)|$ then increases monotonically with $a$, as does the tripartite entanglement of the considered states $\sig^s$ \cite{network,3mpra,renyi}, and saturates to the asymptotic value $|{\cal S}_{\max}^{\infty}| = 16/3^{\frac98} \approx 4.65$ in the limit  $a \rightarrow \infty$, when the state approaches the continuous variable GHZ state. We note that this achievable violation (generalizing the bipartite findings of Ref.~\cite{jeong}), is below the ultimate limit compatible with quantum mechanics, given by $4\sqrt2 \approx 5.65$. We anticipate that  $|{\cal S}_{\max}^{\infty}|$ is the absolute maximum Svetlichny parameter which can be reached by {\it any} three-mode Gaussian state via displaced parity measurements.

We investigate now the Svetlichny violation for arbitrary nonsymmetric pure Gaussian states with three-mode CM $\sig$, Eq.~(\ref{sigma}). One can show that the maximum of Eq.~(\ref{sveltpar}) is attained for measurement settings $\underline \xi_j=(0,p_j)$ and $\underline \xi'_j=(0,-p_j)$. The remaining optimization over $\{p_j\}$ can be solved numerically for each given CM parameterized by $\{a_j\}$ ($j=1,2,3$). It is interesting to analyze in detail how the maximum Svetlichny parameter $|{\cal S}_{\max}\big(\sig(a_1,a_2,a_3)\big)|$  compares with the genuine tripartite entanglement of the states as measured by the residual R\'enyi-$2$ entropy ${\cal E}_{1|2|3}\big(\sig(a_1,a_2,a_3)\big)$ \cite{renyi}. We first observe that $|{\cal S}_{\max}|$ and ${\cal E}_{1|2|3}$ have quite similar fingernail-shaped profiles as functions of the local invariants $a_j$ [Fig.~\ref{figunghie}(a),(b)]. By running an extensive numerical investigation, we find that the maximum Svetlichny parameter admits tight upper and lower bounds as a function of the tripartite entanglement [Fig.~\ref{figunghie}(c)]. The lower bound is saturated by the fully symmetric states studied above, whose $|{\cal S}_{\max}|$ is given by Eq.~(\ref{sveltsym}), and whose tripartite entanglement is ${\cal E}_{1|2|3}\big(\sig(a)\big) = \ln\left[8a^3/\big(4 (a^4+a^2)-f_a(a^2-1)\big)\right]$ \cite{renyi}. The upper bound can be reached by bisymmetric states with $a_1=a_2=a$ and $a_3$ to be found numerically  close to the boundary $a_3\leq 2a-1$.
The analysis in Fig.~\ref{figunghie}(c) shows that
%, while a small portion of fully inseparable states not violating the Svetlichny inequality exists, yet for every nonzero value of ${\cal E}_{1|2|3}$ we can find some (nonsymmetric) pure state which violates inequality (\ref{sveltina}). Furthermore,
all pure three-mode Gaussian states with ${\cal E}_{1|2|3} > \frac12 \ln\big(\frac{32}{27}\big) \approx 0.085$ necessarily violate inequality (\ref{sveltina}). The upper and lower bounds on the Svetlichny parameter close towards $|{\cal S}_{\max}^{\infty}|$ for diverging tripartite entanglement. An example of a nonsymmetric three-mode state which asymptotically reaches $|{\cal S}_{\max}^{\infty}|$ was provided in Ref.~\cite{entropy} for a driven Bose-Einstein condensate in a ring cavity.

We now turn our attention to mixed three-mode Gaussian states.  A numerical investigation of $|{\cal S}_{\max}|$ versus the state purity $\mu$ is presented in Fig.~\ref{figunghie}(d).  In general, the Svetlichny parameter is proportional to $\mu$;
at fixed $\mu$,  $|{\cal S}_{\max}|$ is minimized by product states [which cannot violate inequality (\ref{sveltina})] for which $|{\cal S}^{\rm low}_{\max}|=4\mu$, and is found to admit an upper bound as well, reached by fully symmetric {\it mixed} three-mode Gaussian states, described by a CM $\sig^{ms}(a,\mu)=(\mu)^{-\frac13} \sig^s(a)$.   The upper boundary in Fig.~\ref{figunghie}(d) is specifically obtained in the limit $a \rightarrow \infty$, which gives $|{\cal S}^{\rm up}_{\max}|=\mu |{\cal S}^{\infty}_{\max}|$. Notably, this can fall below $4$ whenever $\mu \leq 3^\frac98/4 \approx 0.86$, which means that no mixed Gaussian state with purity below this threshold can ever violate the Svetlichny inequality.
%This can explain why the thermal three-mode states of the ionic system in Ref.~\cite{jie} were found not to display tripartite nonlocality.

The present study allows us to add one more layer to the diagram characterizing inseparability versus mixedness for symmetric three-mode Gaussian states (first investigated in Ref.~\cite{3mnjp}); see Fig.~\ref{figunghie}(e), which can be read as follows. From bottom-right to top-left: The unfilled region contains fully separable states. The tiny strip between the solid (red) line and the dotted (blue) line accommodates three-mode biseparable states, which exhibit tripartite bound entanglement \cite{giedke,3mnjp}. All the states above the dotted line are fully inseparable. The subregion above the dashed line contains states whose tripartite entanglement has a promiscuous sharing structure \cite{contangle,3mnjp}, namely the reduced bipartite entanglements in all two-mode partitions are nonzero and enhance the genuine tripartite one. Eventually, in the top-left corner, we find the subset of fully inseparable states which display genuine tripartite Svetlichny nonlocality; the shading in such region reproduces the value of $|{\cal S}_{\max}|$, ranging from $4$ (darker) to  $|{\cal S}^\infty_{\max}| \approx 4.65$ (lighter). We recall that the states classified in Fig.~\ref{figunghie}(e), which can be generated e.g.~by letting pure states $\sig^s(a)$ evolve in Gaussian noisy channels \cite{3mpra}, or  by mixing three thermal squeezed beams at a three-mode beam splitter (tritter) \cite{network,3mnjp}, are useful resources for teleportation networks \cite{network,naturusawa}, Byzantine agreement \cite{byz} or quantum secret sharing \cite{secret}.

\begin{figure}[t]
\includegraphics[width=7cm]{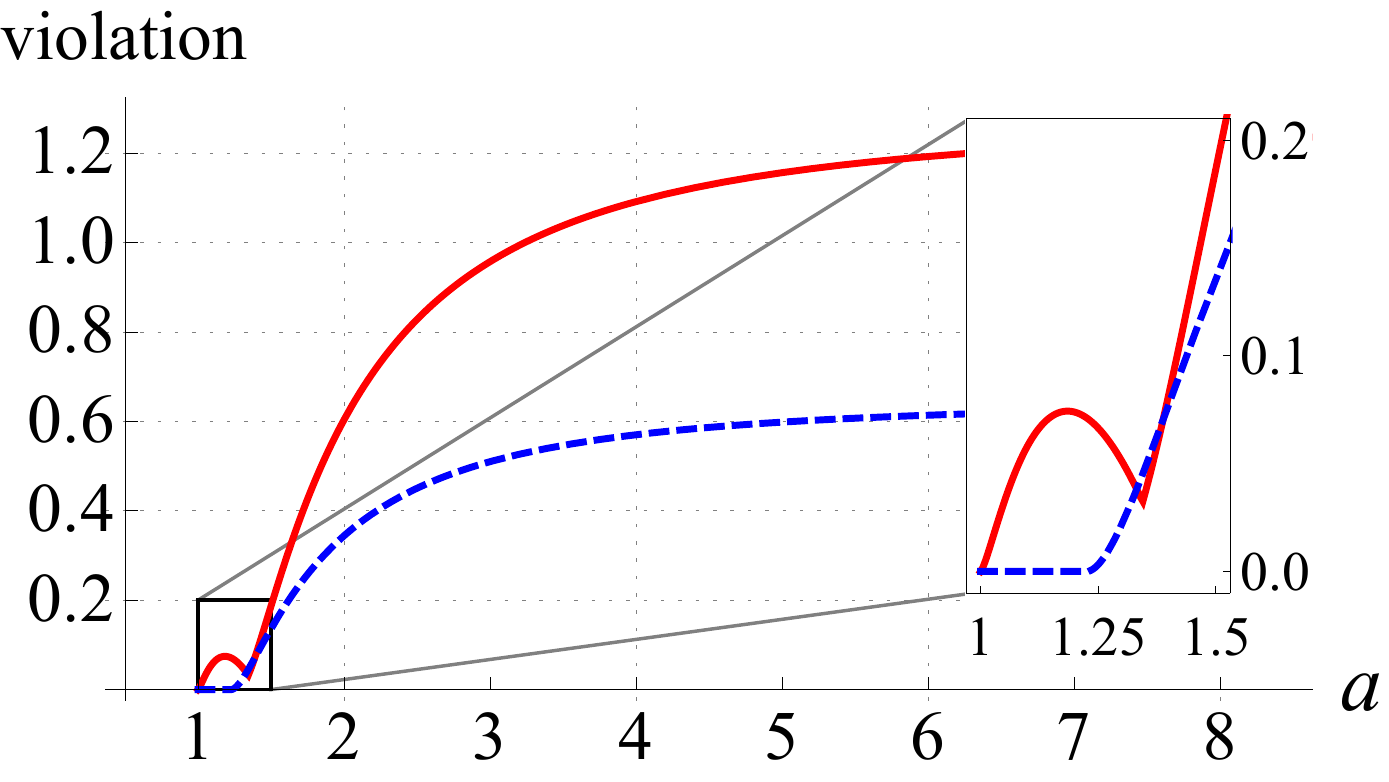}
\caption{Maximum violation of $NS_2$ inequalities $\#168$ (solid red) and $\#185$ (dashed blue, equivalent to Svetlichny inequality (\ref{sveltina}) \cite{notepend}) as defined in \cite{bancal}, plotted for pure three-mode fully symmetric Gaussian states with CM $\sig^s(a)$ versus the local mixedness parameter $a$. The inset displays a zoom into the region $1 \leq a \leq 1.5$. All the quantities plotted are dimensionless.
}
\label{figns2}
\end{figure}

\noindent{\bf Alternative tests of Gaussian three-way nonlocality.}---
Finally, we put to test recently developed criteria to reveal so-called tripartite $NS_2$ nonlocality \cite{bancal}. This analysis reveals that a larger class of Gaussian states can exhibit three-way nonlocality even when inequality (\ref{sveltina}) is not violated. A model is said to be $NS_2$ local if the correlations between any two parties can be simulated by no-signalling (NS) bipartite resources. The constraints imposed by $NS_2$ locality take into account an inherent timing of the local measurements (assumed instead as simultaneous in the Svetlichny analysis \cite{svelt}), which excludes the possibility that outcomes on one or two parties can be used to determine measurement choices on the remaining one, violating causality \cite{bancal}. Consequently, $NS_2$ nonlocality is a three-way form of nonlocality generally weaker than the Svetlichny one. To reveal $NS_2$ nonlocality one needs to observe a nonzero violation in any one of a set of $185$ inequalities \cite{notepend} reported explicitly in the appendix of Ref.~\cite{bancal}. For a Gaussian state, we can reformulate them all in terms of CMs of the state and its marginals, adopting once more displaced parity measurement operators $P_j$ and using their relation to Wigner functions.

%Unlike the Svetlichny inequality, the other $NS_2$ inequalities typically involve also measurements performed by only one or two parties; in this case, we need to consider the Wigner function of the marginal state of the corresponding one-mode or two-mode subsystems. This is all manageable with Gaussian states at the level of CMs.
We have tested $10^5$ random pure three-mode Gaussian states, with particular emphasis on states with $a_j \lesssim 1.5$, including those not violating inequality (\ref{sveltina}). Remarkably, we have found violations of at least one $NS_2$ inequality for all tested fully inseparable pure states. As an example, in Fig.~\ref{figns2} we investigate the violation of  $NS_2$ inequality  $\# 168$ (in the notation of \cite{bancal}) for fully symmetric pure states with CM $\sig^s(a)$, demonstrating that it is violated for all $a>1$, unlike the Svetlicnhy inequality, which is equivalent to the $NS_2$ inequality $\# 185$ \cite{notepend}. We remark that these fully symmetric Gaussian states simultaneously embody normally competing characteristics, typical of both GHZ and $W$ states \cite{contangle,3mpra}; in particular,  $W$-like traits (significant pairwise entanglement in all two-mode reductions) dominate at low squeezing $(a \lesssim 1.5)$ while  GHZ-like traits (maximum genuine tripartite entanglement) dominate for $a \gg 1$. Consequently, different inequalities in the $NS_2$ set exhibit violations in different regimes \cite{notepend}. This explains the nonmonotonic behaviour observed in Fig.~\ref{figns2} (which results from the competition between the two forms of distributed entanglement), and the failure of the Svetlichny inequality (which is best suited to detect GHZ-type entanglement \cite{sveltghzw}) to be violated for small $a$, see Eq.~(\ref{sveltsym}).
%Typically violated inequalities were, in the notation of Ref.~\cite{bancal}, inequality 2, 3, 6, 168, etc.
%We did not seek for the maximum violation across all possible inequalities and settings, just stopped on each state as soon as a violation was detected. Violations as small as $10^{-7}$ were observed.
Our analysis reinforces the conjecture, raised in \cite{bancal} based on a numerical test for three-qubit pure states, that {\it all} fully inseparable pure tripartite states in arbitrary dimensions are in fact three-way nonlocal, mirroring the seminal result that all pure bipartite entangled states violate a Bell-CHSH inequality \cite{gisin}.

%\begin{figure}[h]

\noindent{\bf Conclusions.}---
We have performed the first systematic study of three-way nonlocality for Gaussian states of continuous variable systems, providing unambiguous evidence for their nonclassical nature, and revealing close connections with genuine tripartite entanglement.  Our findings yield practical recipes for the demonstration of multipartite continuous variable nonlocality in experiments \cite{naturusawa,paulo}, delivering precise prescriptions for the phase space points where maximal violations are detectable. A generalization to $n$-mode states will be the subject of further work.
%A verification of our study appears in reach e.g. in setups involving three-color entanglement \cite{paulo}.

\noindent {\bf Acknowledgments.}--- {We acknowledge fruitful exchanges with A. Ac\'in, J.-D. Bancal, J. Li, L. Mazzola, M. Navascu\'es, and M. Paternostro. We thank the University of Nottingham for financial support via a Nottingham Advance
Research Fellowship (S.P.) and a Nottingham-Tsinghua Fund Grant (G.A.).}


\begin{thebibliography}{99}

\bibitem{nv}
G. Adesso and D. Girolami, Nature Photon. {\bf 6}, 579 (2012).

\bibitem{EPR35}
A.~Einstein, B.~Podolsky, and N.~Rosen, Phys. Rev. \textbf{47}, 777 (1935).


\bibitem{bell}
J.~S. Bell, Physics \textbf{1}, 195 (1964); J.~S. Bell, \emph{Speakable and Unspeakable in Quantum Mechanics}\ (Cambridge
  University Press, Cambridge, 1987).


\bibitem{chsh}J.~Clauser, M.~Horne, A.~Shimony, and R.~Holt, Phys. Rev. Lett. \textbf{23},
  880 (1969).

\bibitem{bellrev}
N. Brunner, D. Cavalcanti, S. Pironio, V. Scarani, and S. Wehner, arXiv:1303.2849 (2013).


\bibitem{entanglement}
R. Horodecki,  P. Horodecki, M. Horodecki, and K. Horodecki, Rev. Mod. Phys. {\bf 81}, 865 (2009).

\bibitem{schr}
E.~Schr{\"o}dinger, Naturwiss. \textbf{23}, 812 (1935).


\bibitem{werner89}
R.~F. Werner, Phys. Rev. A \textbf{40}, 4277 (1989).

\bibitem{operationonloc}
R. Gallego, L. E. W\"urflinger, A. Ac\'in, and M. Navascu\'es,  Phys. Rev. Lett. {\bf 109}, 070401 (2012).

\bibitem{qkd}
J. Barrett, L. Hardy, and A. Kent, Phys. Rev. Lett. {\bf 95},
010503 (2005);  A. Ac\'in, N. Brunner, N. Gisin, S. Massar, S. Pironio,
and V. Scarani, Phys. Rev. Lett. {\bf 98}, 230501 (2007); L. Masanes, Phys. Rev. Lett. {\bf 102}, 140501 (2009);
L. Masanes, S. Pironio, and A. Ac\'in, Nature Commun. {\bf 2}, 238 (2011).





\bibitem{merminklyshko}
 N. D. Mermin, Phys. Rev. Lett. {\bf 65}, 1838 (1990); D. N. Klyshko,
Phys. Lett. A {\bf 172}, 399 (1993).

\bibitem{svelt}
G. Svetlichny, Phys. Rev. D {\bf 35}, 3066 (1987).

\bibitem{bancalprl}
 J.-D. Bancal, N. Brunner, N. Gisin, and Y.-C. Liang, Phys. Rev.
Lett. {\bf 106}, 020405 (2011).


\bibitem{sveltghzw}
S. Ghose, N. Sinclair, S. Debnath, P. Rungta, and R. Stock, Phys. Rev. Lett. {\bf 102}, 250404 (2009);
A. Ajoy and P. Rungta, Phys. Rev. A {\bf 81}, 052334 (2010).


\bibitem{expsvelt}
 J. Lavoie, R. Kaltenbaek, and K. J. Resch, New J. Phys. {\bf 11},
073051 (2009).


\bibitem{othermulti}
M. Seevinck and G. Svetlichny, Phys. Rev. Lett. {\bf 89}
060401 (2002); D. Collins, N. Gisin, S. Popescu, D. Roberts, and
V. Scarani, Phys. Rev. Lett. {\bf 88}, 170405 (2002);
P. Mitchell, S. Popescu and D. Roberts, Phys. Rev. A {\bf 70},
060101 (2004); N. S. Jones, N. Linden, and S. Massar, Phys. Rev. A {\bf 71},
042329 (2005); J.-D. Bancal, C. Branciard, N. Gisin, and
S. Pironio, Phys. Rev. Lett. {\bf 103}, 090503 (2009).


\bibitem{bancal}
J.-D. Bancal, J. Barrett, N. Gisin, and S. Pironio, arXiv:1112.2626v2, Phys. Rev. A (2013), in press.

 \bibitem{ckw}
 V. Coffman, J. Kundu, and W. K. Wootters, Phys. Rev. A {\bf 61}, 052306 (2000).


\bibitem{homobell}
%W. J. Munro, Phys. Rev. A {\bf 59}, 4197 (1999);
H. Nha and H. J.
Carmichael, Phys. Rev. Lett. {\bf 93}, 020401 (2004);  R.
Garc\'ia-Patr\'on,  J. Fiur\'a{\u s}ek, N. J. Cerf, J. Wenger, R. Tualle-Brouri, and
P. Grangier, Phys. Rev. Lett. {\bf 93}, 130409
(2004); R. Garc\'ia-Patr\'on, J. Fiur\'a{\u s}ek, and N. J. Cerf,  Phys. Rev. A {\bf 71}, 022105
(2005).

\bibitem{kb} K. Banaszek and K. Wodkiewicz, Phys. Rev. Lett. {\bf 76}, 4344
(1996).

\bibitem{jeong}
H. Jeong, W. Son, M. S. Kim, D. Ahn, and C. Brukner, Phys. Rev. A {\bf 67}, 012106 (2003).


\bibitem{bramermin}
P. van Loock and S. L. Braunstein, Phys. Rev. A {\bf 63},   022106 (2001).

\bibitem{chen}
Z. B. Chen, J. W. Pan, G. Hou, and Y. D. Zhang, Phys Rev Lett. {\bf 88}, 040406 (2002).
\bibitem{ferraroparis}
 A. Ferraro and M. G. A. Paris, J. Opt. B: Quantum Semiclass.
Opt. {\bf 7}, 174 (2005).

\bibitem{maurorefs}
 A. Ac\'in, N. J. Cerf, A. Ferraro, and J. Niset, Phys. Rev. A {\bf 79},
012112 (2009).

\bibitem{jie}
J. Li, T. Fogarty, C. Cormick, J. Goold, T. Busch, and
M. Paternostro, Phys. Rev. A {\bf 84}, 022321 (2011).

\bibitem{mauro}
S. W. Lee, M. Paternostro, J. Lee, and H. Jeong, Phys. Rev. A {\bf 87}, 022123 (2013).


 \bibitem{brareview}
 S. L. Braunstein and P. van Loock, Rev. Mod. Phys. {\bf 77}, 513
(2005).

  \bibitem{gaussreview}
C. Weedbrook, S. Pirandola, R. Garcia-Patron, N. J. Cerf, T. C. Ralph, J. H. Shapiro, and S. Lloyd,
Rev. Mod. Phys. {\bf 84}, 671 (2012).

 \bibitem{ourreview}
 G. Adesso and F. Illuminati, J. Phys. A: Math. Theor. {\bf 40} 7821, (2007).


 \bibitem{book}
 N.~Cerf, G.~Leuchs, and E.~S. Polzik (eds.), \emph{Quantum
Information with
  Continuous Variables of Atoms and Light}\ (Imperial College Press, London,
  2007).


\bibitem{nogo}
 J. Fiur\'a\v{s}ek,  Phys. Rev. Lett. {\bf 89}, 137904 (2002);
J.~Eisert, S.~Scheel, and M.~B. Plenio, Phys. Rev. Lett. \textbf{89}, 137903
  (2002);
 G. Giedke and J. I. Cirac, Phys. Rev. A {\bf 66}, 032316 (2002).


 \bibitem{renyi}
G. Adesso, D. Girolami, and A. Serafini, Phys. Rev. Lett. {\bf 109}, 190502 (2012).


\bibitem{nonclasswig}
A. Mari, K. Kieling, B. Melholt Nielsen, E. S. Polzik, and J. Eisert,
Phys. Rev. Lett. {\bf 106}, 010403 (2011).


\bibitem{gaussimul}
A. Mari and J. Eisert, Phys. Rev. Lett. {\bf 109}, 230503 (2012).


\bibitem{entropy}
S. Piano and G. Adesso, Entropy {\bf 15}, 1875 (2013).


\bibitem{3mpra}
 G. Adesso, A. Serafini, and F. Illuminati, Phys. Rev. A {\bf 73}, 032345 (2006).

\bibitem{3mnjp}
G. Adesso, A. Serafini, and F. Illuminati, New J. Phys. {\bf 9}, 60 (2007).


 \bibitem{contangle}
 G. Adesso and F. Illuminati, New J. Phys. {\bf 8}, 15 (2006).

\bibitem{giedke}
G.~Giedke, B.~Kraus, M.~Lewenstein, and J.~I. Cirac, Phys. Rev. A \textbf{64},
  052303 (2001).

\bibitem{network}
P.~{van Loock} and S.~L. Braunstein, Phys. Rev. Lett. \textbf{84}, 3482 (2000); G. Adesso and F. Illuminati,
 Phys. Rev. Lett.
  \textbf{95}, 150503 (2005).


\bibitem{naturusawa}
H.~Yonezawa, T.~Aoki, and A.~Furusawa, Nature \textbf{431}, 430 (2004).

\bibitem{byz}
R. Neigovzen, C. Rod\'o, G. Adesso, and A. Sanpera,
Phys. Rev. A {\bf 77}, 062307 (2008).


\bibitem{secret}
A.~M. Lance, T.~Symul, W.~P. Bowen, B.~C. Sanders, and P.~K. Lam, Phys. Rev.
  Lett. \textbf{92}, 177903 (2004).

\bibitem{notepend}
 Precisely, there are $185 \times 2^6$ inequalities, as for each one there are two possible relabellings for the outcomes of each local measurement. The $\# 185$ $NS_2$ inequality facet in \cite{bancal} is the conventional Svetlichny one \cite{svelt}. Notice that each of those inequalities can be violated up to a different extent, so the quantitative amounts of their violation are incomparable. Some inequalities such as $\# 1$ can never be violated. Others can be violated only for certain types of fully inseparable states, or certain ranges of local CM parameters in the case of Gaussian states.

\bibitem{gisin}
N.~Gisin, Phys. Lett. A \textbf{154}, 201 (1991).

\bibitem{paulo}
A. S. Coelho, F. A. S. Barbosa, K. N. Cassemiro, A. S. Villar, M. Martinelli, and P. Nussenzveig, Science {\bf 326}, 823 (2009).

\bibitem{rmp}
H.  Ritsch, P. Domokos, F. Brennecke, and T. Esslinger, Rev. Mod. Phys. {\bf 85}, 553 (2013).






\end{thebibliography}
\end{document}